\documentclass[aps,prl,showpacs,twocolumn]{revtex4}
\usepackage{epsfig}
\begin{document}

\title
{Quantum state filtering and discrimination between sets of Boolean
  functions} 
\author{J\'{a}nos A. Bergou$^{1,2}$}
\author{Ulrike Herzog$^3$}
\author{Mark Hillery$^1$}
\affiliation{$^1$Department of Physics, Hunter College, City
    University of New York, 695 Park Avenue, New York, NY 10021,
    USA}
\affiliation{$^2$Institute of Physics, Janus Pannonius University,
    H-7624 P\'{e}cs, Ifj\'{u}s\'{a}g \'{u}tja 6, Hungary}
\affiliation{$^3$Institut f\"ur Physik,  Humboldt-Universit\"at zu
    Berlin, Newtonstrasse 15,  D-12489 Berlin, Germany}

\date{\today}
\begin{abstract}
In quantum state filtering one wants to determine
whether an unknown quantum state, which is chosen from a known set of
states, $\{|\psi_{1}\rangle,\ldots |\psi_{N}\rangle\}$, is either a
specific state, say $|\psi_1\rangle$, or one of the remaining states,
$\{|\psi_{2}\rangle,\ldots |\psi_{N}\rangle\}$. We present the 
optimal solution to this problem, in terms of generalized
measurements, for the case that the filtering is required to
be unambiguous. As an application, we propose an efficient,
probabilistic quantum algorithm for distinguishing between sets of
Boolean functions, which is a generalization of the Deutsch-Jozsa
algorithm.
\end{abstract}


\maketitle

Optimal discrimination among quantum states plays a central role in
quantum information theory. Interest in this problem was prompted by
the suggestion to use 
nonorthogonal quantum states for communication in certain secure
quantum cryptographic protocols, most notably in the one based on the
two-state procedure as developed by Bennett \cite{bennett}. The
reason why until recently the area has shown relatively slow progress
within the rapidly evolving field of quantum information is that it
poses quite formidable mathematical challenges. Except for a
handful of very special cases, no general exact solution has been
available involving more than two arbitrary states. In this paper we
present an exact solution to an optimum
measurement problem involving an {\it arbitrary} number of quantum
states, with {\it no restriction} on the states. The resulting 
method has the potential for widespread applications in quantum
information processing. In particular, it lends itself quite naturally
to a quantum generalization of probabilistic classical
algorithms. Whenever it is possible to find a one-to-one 
mapping of classical alternatives onto quantum states, our method can
discriminate among these quantum 
alternatives in a {\it single step} with optimum success probability. 

We illustrate the strength of the method on the example of a
probabilistic quantum algorithm to 
discriminate between sets of Boolean functions. A Boolean function on
$n$ bits is one that returns either 0 or 1 as output for every
possible value of the input $x$, where $0\leq x\leq 2^{n}-1$.  The
function is uniform (or constant) if it returns the same output on all
of its arguments, i.e.\ either all 0's or all 1's; it is balanced (or
even) if it returns 0's on half of its arguments and 1's on the other
half; and it is biased if it returns 0's on $m_0$ of its arguments and
1's on the remaining $m_1=2^{n}-m_{0}$ arguments ($m_{0}\neq m_{1}\neq
0$ or $2^{n}-1$). Classically, if one is given an unknown function and
told that it is either balanced or uniform, one needs
$2^{(n-1)}+1$ measurements to decide which. Deutsch and Jozsa
\cite{DJ} developed a quantum algorithm that can accomplish this
task in one step. To discriminate a biased Boolean function from
an unknown balanced one, $2^{(n-1)}+m_{1}+1$ measurements are
needed classically, where, without loss of generality, we have
assumed that $m_{1}<m_{0}$. Here  we propose a probabilistic
quantum algorithm that can unambiguously discriminate a known
biased Boolean function from a given set of balanced ones in one
step.

The method is based on the optimum unambiguous quantum state filtering
scenario which, in turn, is a
special case of the following more general problem. We know that a
given system is prepared in one of $N$ known non-orthogonal
quantum states, but we do not know which one. We want to assign
the state of this system to one or the other of two
complementary subsets of the set of the $N$ given states where
one subset has $M$ elements and the other has $N-M$ ($M \leq
N/2$). Since the subsets are not mutually orthogonal, the assignment
can not be done with a 100\% probability of success. For the case 
that the assignment is required to be unambiguous, at the expense
of allowing inconclusive results to occur the probability of which
is minimized, the problem has recently been solved for $N=3$
\cite{sun2}.  For the case that the assignment is to be performed
with minimum error, the solution has been found for arbitrary $M$
and $N$ under the restriction that the Hilbert space spanned by
the states is two-dimensional \cite{HB}. We refer to either case as
quantum state filtering when $M=1$ and $N \geq 3$.

Unambiguous filtering is related to unambiguous quantum state
discrimination: one is given a quantum
system, prepared in a state that belongs to a known set of
non-orthogonal states, and one wants to determine, without possibility
of error, which state the system is in \cite{chefrev}. Since the
states are not mutually orthogonal, at first glance the problem
appears impossible. However, it becomes possible if we allow the
procedure to fail a certain fraction of the time. That is, when we
apply the procedure, we will either find out what the quantum state of
the system is, or we will fail to do so, but we will never make an
erroneous identification. The optimal method for discriminating
between two states was found in 
Refs. \cite{ivanovic}-\cite{peres}.  No general solution is known for
more than two states but there are special cases that can be solved,
and some features of the general solution have been extracted
\cite{terno}-\cite{sun1}. Chefles has shown \cite{chefles1} that
the states have to be linearly independent for unambiguous state
discrimination to succeed but orthogonality is not required.

We begin by presenting the solution to the unambiguous quantum
state filtering problem.  Suppose we are given a quantum system
prepared in the state $|\psi\rangle$, which is guaranteed to be a
member of the set of $N$ non-orthogonal states
$\left\{|\psi_{1}\rangle,\ldots,|\psi_{N}\rangle\right\}$, but we
do not know which one.  We denote by $\eta_i$ the {\em a priori}
probability that the system was prepared in the state
$|\psi_{i}\rangle$.  We want to find a procedure that will
unambiguously assign the state of the quantum system to one or the
other of two complementary subsets of the set of the $N$ given
non-orthogonal quantum states, either $\{ |\psi_{1}\rangle\}$ or
$\{ |\psi_{2}\rangle ,\ldots |\psi_{N}\rangle \}$.  Quantum
measurement theory tells us that non-orthogonal states cannot be
discriminated perfectly.  If we are given $|\psi_{i}\rangle$, we will
have some probability $p_{i}$ to correctly assign it to one of the
subsets and, correspondingly, some failure probability, $q_{i}=
1-p_{i}$, to obtain an inconclusive answer. The average probabilities
of success and of failure are
    \label{P}
 $   P = \sum_{i=1}^{N}{\eta}_{i}p_{i}$,
and
\begin{equation}
    \label{Q}
    Q = \sum_{i=1}^{N}{\eta}_{i}q_{i} ,
\end{equation}
respectively. Our objective is to find the set of $\left\{ q_i
\right\}$ that minimizes the probability of failure, $Q$.

It is easy to see that a standard quantum measurement (SQM, a von
Neumann projective measurement) can achieve error-free filtering. If
we project on either of the two sets (state selective measurement,
first strategy) a ``no click'' will indicate that we were given a
state from the
other set, assuming perfect detectors. A somewhat better approach is
to project on a direction that is perpendicular to one of the sets
(nonselective measurement, second strategy). Now, a detector ``click''
indicates that we were given a state from the other set and
perfect detectors are not required. For example, if we measure the
operator $F^{(1)}=I-|\psi_{1}\rangle\langle\psi_{1}|$, then a click
(corresponding to the eigenvalue $1$) shows that the vector is not
$|\psi_{1}\rangle$ and the measurement has succeeded.  If we do not
obtain a click (eigenvalue $0$), then the measurement has failed, and
we do not know which vector we were given.  The probability of
failure, $Q_{SQM}^{(1)}$, is given by
\begin{equation}
    Q_{SQM}^{(1)}=\eta_{1} + S  ,
    \label{Qsqm1}
\end{equation}
where
$S=\sum_{i=2}^{N}\eta_{i}|\langle\psi_{1}|\psi_{i}\rangle|^{2}$
is the average overlap between the two subsets.

A second possibility is to split $|\psi_{1}\rangle$ into two
components, $|\psi_{1}\rangle = |\psi_{1}^{\perp}\rangle +
|\psi_{1}^{\parallel}\rangle$.  Here $|\psi_{1}^{\perp}\rangle$ is
orthogonal to the subspace, $\mathcal{H}_{2}$, that is spanned by the
vectors $|\psi_{2}\rangle, \ldots |\psi_{N}\rangle$, and
$|\psi_{1}^{\parallel}\rangle$ lies in $\mathcal{H}_{2}$. Their 
normalized versions are $|\tilde{\psi}_{1}^{\perp}\rangle =
|\psi_{1}^{\perp}\rangle / \|\psi_{1}^{\perp}\|$ and
$|\tilde{\psi}_{1}^{\parallel}\rangle = |\psi_{1}^{\parallel}\rangle /
\|\psi_{1}^{\parallel}\|$, respectively, where the norm is defined in
the usual way, $\| \psi \|^{2} = \langle \psi | \psi \rangle$. We
then introduce the operator
$F^{(2)}=|\tilde{\psi}_{1}^{\perp}\rangle\langle
\tilde{\psi}_{1}^{\perp}| -
(I - |\tilde{\psi}_{1}^{\perp}\rangle\langle\tilde{\psi}_{1}^{\perp}|
-|\tilde{\psi}_{1}^{\parallel}\rangle\langle
\tilde{\psi}_{1}^{\parallel}|)$,
which has eigenvalues $1$, $0$, and $-1$.  If we measure $F^{(2)}$ and
obtain $1$, then 
the vector was $|\psi_{1}\rangle$, if we obtain $-1$, then the vector
was in the set $\{ |\psi_{2}\rangle , \dots |\psi_{N}\rangle\}$, and
if we obtain $0$, the procedure failed. In this case the probability
of failure, $Q_{SQM}^{(2)}$, is given by
\begin{equation}
\label{Qsqm2}
Q_{SQM}^{(2)}=\eta_{1}\|\psi_{1}^{\parallel}\|^{2} +
\frac{S}{\|\psi_{1}^{\parallel}\|^{2}}.
\end{equation}
Which of these two particular strategies is better is determined by
which of these two failure probabilities is smaller. In particular,
$Q_{SQM}^{(1)}>Q_{SQM}^{(2)}$ if $\eta_{1}\|\psi_{1}^{\parallel}\|^{2}
>S$, and vice versa.

Now, the question arises: Is this the best we can do? The answer
is that under certain conditions a generalized measurement based
on positive-operator valued measures (POVM, \cite{kraus}) can do
better in an intermediate range of parameters, and can achieve a
higher probability of success than a standard von Neumann
measurement. The POVM can be implemented by a unitary evolution on
a larger space and a selective measurement. The larger space
consists of two orthogonal subspaces, the original system space and a
failure space. The unitary evolution transforms
the input sets into orthogonal sets in the original system space
and maps them onto the same vector in the failure space.  A click in
the detector measuring along this vector
corresponds to failure of the procedure, since all inputs are
mapped onto the same output. A no-click corresponds to success
since now the non-orthogonal input sets are transformed into
orthogonal output sets in the system space.  The
one-dimensionality of the failure space follows from the
requirement that the filtering is optimum. Namely, suppose that
$|\psi_{1}\rangle$ is mapped onto some vector in the failure space
and the inputs from the other set are mapped onto vectors that
have components perpendicular to this vector.  Then a single von
Neumann measurement along the orthogonal direction could identify
the input as being from the second set, i.e. further filtering
would be possible, lowering the failure probability and the original
filtering could not have been optimum.

In particular, let ${\cal H}_{S}$ be the $D$-dimensional system
space spanned by the vectors $\{ |\psi_{1}\rangle ,\ldots
|\psi_{N}\rangle\}$ where, obviously, $D \leq N$.  We now embed
this space in a space of $D+1$ dimensions, ${\cal H}_{S+A}={\cal
H}_{S} \oplus {\cal H}_{A}$, where ${\cal H}_{A}$ is a
one-dimensional auxiliary Hilbert space, the failure space or ancilla.
The basis in this space is denoted by $|\phi^{(A)}\rangle$.  Thus, the
unitary evolution on ${\cal H}_{S+A}$ is specified by the requirement
that for any input state $|\psi_{i}\rangle
(=|\psi_{i}^{(S)}\rangle)$ ($i=1,\ldots,N$) the final state has the
structure
\begin{equation}
|\psi_{i}\rangle_{out}=U|\psi_{i}\rangle = \sqrt{p_{i}}
 |\psi_{i}^{\prime (S)}\rangle + \sqrt{q_{i}} e^{i\theta_{i}}
|\phi^{(A)}\rangle \  .
\label{U}
\end{equation}
From unitarity  the relation, $p_{i}+q_{i}=1$, follows.
Furthermore, $p_{i}$ is the probability that the transformation
$|\psi_{i}\rangle \rightarrow |\psi^{\prime}_{i}\rangle$ succeeds
and $q_{i}$ is the probability that $|\psi_{i}\rangle$ is mapped
onto the state $|\phi^{(A)}\rangle$.  In order  to identify $p_{i}$
and $q_{i}$ with the state-specific success and failure
probability for quantum filtering we have to require that
\begin{equation}
\langle\psi^{\prime}_{1}| \psi^{\prime}_{i}\rangle = 0 ,
\label{ortho}
\end{equation}
for $i=2,\ldots N$.  We now introduce the operator 
$F^{(3)}=|\psi_{1}^{\prime}\rangle\langle \psi_{1}^{\prime}| -
(I^{(S+A)} - |\psi_{1}^{\prime}\rangle\langle \psi_{1}^{\prime}|
-|\phi^{(A)}\rangle\langle \phi^{(A)}|)$,
which has eigenvalues $1$, $0$, and $-1$.  If we measure $F^{(3)}$ and
obtain $1$, then the input was $|\psi_{1}\rangle$, if we obtain $-1$,
then the input was from the other the set, and if we obtain $0$, the
procedure failed. 

In order to optimize the POVM, we have to determine those values of
$q_{i}$ in Eq. (\ref{U}) that yield the smallest average failure
probability $Q$.  Taking the scalar product of $U|\psi_{1}\rangle$
and $U|\psi_{i}\rangle$ in Eq. (\ref{U}), and using Eq.
(\ref{ortho}), gives
\begin{equation}
|\langle \psi_{1}|\psi_{i}\rangle|^{2}=q_{1} q_{i} ,
\label{qproduct}
\end{equation}
for $i=2,\ldots,N$, and Eq. (\ref{Q}) can be cast in the form
$Q(q_{1}) = \eta_{1} q_{1} +S/q_{1}$. Unitarity of the
transformation $U$ delivers the necessary condition that
$q_{1}$ must lie in the range $\|\psi_{1}^{\parallel}\|^{2} \leq
q_{1} \leq 1$. Details of the derivation, along with a discussion of
the sufficient conditions for the existence of $U$, will be presented
in a future publication \cite{BHH}. Provided that a POVM-solution
exists, the minimum of $Q(q_{1})$ is reached for $q_{1} =
\sqrt{S/\eta_{1}}$ and is given by
\begin{equation}
Q_{POVM} = 2 \sqrt{\eta_{1} S}.
\label{Qpovm}
\end{equation}

\begin{figure}[ht]
\epsfig{file=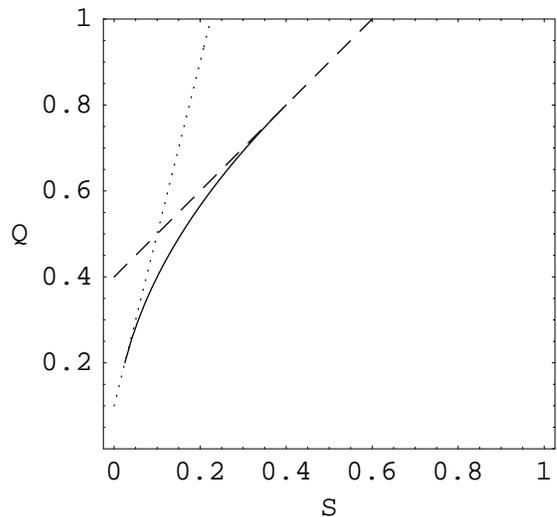, height=7cm}
\caption{Failure probability, $Q$, vs. the average overlap,
  $S$. Dashed line: $Q_{SQM}^{(1)}$, dotted line: $Q_{SQM}^{(2)}$,
  solid line: $Q _{POVM}$. For the figure we used the
  following representative values: $\eta_{1}=0.4$ and
   $\|\psi_{1}^{\parallel}\|^{2}=0.25$. For these the optimal
  $Q$ is given by $Q_{SQM}^{(2)}$ for $0<S<0.025$, by
  $Q_{POVM}$ for $0.025 \leq S \leq
  0.4$ and by $Q_{SQM}^{(1)}$ for $0.4<S$.}
\label{Fig1}
\end{figure}

Thus, the failure probability for optimal unambiguous quantum
state filtering can be summarized as
\begin{eqnarray}
    \label{Qmin}
    Q = \left\{ \begin{array}{ll}
    2 \sqrt{\eta_{1} S}
    & \mbox{if
    $\eta_{1}\| \psi_{1}^{\parallel} \|^{4}
    \leq  S
    \leq \eta_{1}$} , \\
    \eta_{1}+S & \mbox{ if $S > \eta_{1}$} , \\
    \eta_{1}\|\psi_{1}^{\parallel}\|^{2}  + \frac{S}
    {\|\psi_{1}^{\parallel}\|^{2}} & \mbox{ if $S
    < \eta_{1} \|\psi_{1}^{\parallel} \|^{4}$} .
    \end{array}
    \right.
\end{eqnarray}
The first line represents the POVM result, Eq.\ (\ref{Qpovm}), and it
gives a smaller failure probability, in its range of validity, than the
von Neumann measurements, Eqs.\ (\ref{Qsqm1}) and Eq.\
(\ref{Qsqm2}), cf.  Fig. \ref{Fig1}. Outside of the POVM range
of validity we recover the von Neumann results. It should be noted
that for these results to hold, unlike for unambiguous state
discrimination, linear independence of all states is not
required. Instead, the less stringent requirement of the linear
independence of the sets is sufficient, in agreement with the findings
in \cite{zhang}.

We can now apply this result to
distinguishing between sets of Boolean functions.  Let $f(x)$,
where $0\leq x\leq 2^{n}-1$, be a Boolean function, i.e.\ $f(x)$
is either $0$ or $1$.  One of the sets we want to consider is a
set of balanced functions.  In our example, the second set has
only two members, and we shall call it ${\cal W}_{k}$. A function is
in ${\cal W}_{k}$ if $f(x)=0$ for $0\leq x< [(2^{k}-1)/2^{k}]2^{n}$
and $f(x)=1$ for $[(2^{k}-1)/2^{k}]2^{n}\leq x\leq 2^{n}-1$, or if
$f(x)=1$ for $0\leq x< [(2^{k}-1)/2^{k}]2^{n}$ and $f(x)=0$ for
$[(2^{k}-1)/2^{k}] 2^{n}\leq x\leq 2^{n}-1$. We now wish
to distinguish between the given balanced functions
and functions in ${\cal W}_{k}$, that is, we are given an unknown
function that is in one of the two sets, and we want to find out
which set it is in. We note that the two functions in ${\cal W}_{k}$
are biased functions, so that this is a special case of
a more general problem of distinguishing a set of biased functions
from balanced functions.

This is by no means the only example the method can handle, but it is
a particularly simple one and represents a generalization of the
Deutsch-Jozsa problem \cite{DJ}. 
In that case one is given an unknown function that is either
balanced or constant, and one wants to determine which.
Classically, in the worst case one would have to evaluate the
function $D/2+1$ times, where we have set $D=2^{n}$, but in the
quantum case only one evaluation is necessary.  The
solution makes use of the unitary mapping
\begin{equation}
|x\rangle |y\rangle\rightarrow |x\rangle |y+f(x)\rangle ,
\end{equation}
where the first state, $|x\rangle$, is an $n$-qubit state, the
second state, $|y\rangle$, is a single qubit state, and the
addition is modulo $2$.  The state $|x\rangle$, where $x$ is an
$n$-digit binary number, is a member of the computational basis
for $n$ qubits, and the state $|y\rangle$, where $y$ is either $0$
or $1$, is a member of the computational basis for a single qubit.
In solving the Deutsch-Jozsa problem, this mapping is employed in
the following way
\begin{equation}
\label{map2}
\sum_{x=0}^{D-1}|x\rangle (|0\rangle -|1\rangle ) \rightarrow 
\sum_{x=0}^{D-1}(-1)^{f(x)}|x\rangle (|0\rangle - |1\rangle ) .
\end{equation}
This has the effect of mapping Boolean functions to vectors in the
$D$-dimensional Hilbert space, ${\cal H}_{D}$, and we shall do the
same.  The final qubit is not entangled with the remaining $n$
qubits and can be discarded.  The vectors
$\sum_{x=0}^{D-1}(-1)^{f(x)} |x\rangle$ that are produced by
balanced functions are orthogonal to those produced by constant
functions.  This is why the Deutsch-Jozsa problem is easy to solve
quantum mechanically.  In our case, the vectors produced by
functions in ${\cal W}_{k}$ are not orthogonal to those produced by
balanced functions.  However, unambiguous quantum state filtering
provides an optimum probabilistic quantum algorithm for the solution
of this problem.

In order to apply the filtering solution, we note
that both functions in ${\cal W}_{k}$ are mapped, up to an overall sign,
to the same vector in ${\cal H}_{D}$, which we shall call
$|w_{k}\rangle$. The vectors that correspond to balanced functions
are contained in the subspace, ${\cal H}_{b}$, of ${\cal H}_{D}$,
where ${\cal H}_{b}=\{ |v\rangle\in {\cal H}_{D}|\sum_{x=0}^{D-1}
v_{x}=0\}$, and $v_{x}=\langle x|v\rangle$. This subspace has
dimension $2^{n}-1=D-1$, and it is possible to choose an orthonormal
basis, $\{|v_{i}\rangle |i=2,\ldots D\}$, for it in which each basis 
element corresponds to a particular balanced Boolean function 
\cite{BHH}. 

Let us first see how the filtering procedure performs when
applied to the problem of distinguishing
$|w_{k}\rangle(=|\psi_{1}\rangle)$ from the set of the 
$D-1$ orthonormal basis states, $|v_{i}\rangle(=|\psi_{i}\rangle)$, in  
${\cal H}_{b}$. We assume their \emph{a priori} probabilities to be
equal, i.e.\ $\eta_{i}=\eta = (1-\eta_{1})/(D-1)$ for $i=2,\ldots D$,
where $\eta_1$ is the \emph{a priori} probability for
$|w_k\rangle$. For $\|\psi_{1}^{\parallel}\|^{2} =
\|w_{k}^{\parallel}\|^{2} \equiv f_{k}$ we obtain $f_{k} =
(2^{k}-1)/2^{2k-2}$. Then the average overlap, $S_{k}$,
between $|w_{k}\rangle$ and the set of balanced basis vectors can be
written as
\begin{equation}
S_{k} = \frac{1-\eta_{1}}{D-1}f_{k}\ ,
\label{Sk}
\end{equation}
in terms of $f_{k}$ \cite{BHH}.
The failure probabilities are given by Eq.\ (\ref{Qmin}), using
$S=S_{k}$ and, to good approximation, the POVM result holds when
$1/2^{k-2}\leq D \eta_{1}\leq 2^{k-2}$.
For example, in the case in which all of the \emph{a priori}
probabilities are equal, i.e.\ $\eta_{1}=1/D$, we find that
$Q_{SQM}^{(1)} = Q_{SQM}^{(2)} = Q_{SQM} = (1+f_{k})/D$. From Fig.\
\ref{Fig1} the
difference between the POVM and the von Neumann measurement is at its 
largest. To good approximation, $Q_{POVM}/Q_{SQM}=4/2^{k/2}$, which,
for $ k\gg 1$, shows that the POVM can perform substantially better
than the von Neumann measurements.

Now that we know how this procedure performs on the basis vectors
in ${\cal H}_{b}$, we shall examine its performance on any balanced 
function, i.e. we apply it to the problem of distingushing
$|w_{k}\rangle$ from the set of all states in ${\cal H}_{b}$ that
correspond to balanced functions. The number of such states is 
$N=D!/(D/2)!^{2}$ and we again assume their {\it a priori}
probabilities to be equal, $\eta=(1-\eta_{1})/N$. It can be shown 
\cite{BHH} that the average overlap
between $|w_{k}\rangle$ and the set $\{|v\rangle\}$ is given 
by the same expression, Eq.\ (\ref{Sk}), as in the previous
case. Therefore, much of what was said in the previous paragraph
remains valid for this case, as well, with one notable difference. The
case $\eta_{1}=1/D$ now does not correspond to equal {\it a priori} 
probability for the states but, rather, to {\it a priori} weight of
the sets that is proportional to their dimensionality. In this case it
is the POVM that performs best. In the case of equal {\it a priori}
probability for all states, $\eta_{1}=1/(N+1)$, we are outside of the
POVM range of validity and it is the first standard quantum
measurement (SQM1) that performs best. Both the POVM and the SQM1 are
good methods for distinguishing functions in ${\cal W}_{k}$ from
balanced functions. Which one is better depends on the 
\emph{a priori} probabilities of the functions.

Classically, in the worst case, one would have to evaluate a
function $2^{n}[(1/2)+(1/2^{k})]+1$ times to determine if it is in
${\cal W}_{k}$ or if it is an even function.  Using quantum
information processing methods, one has a very good
chance of determining this with only one function evaluation. This
shows that Deutsch-Jozsa-type algorithms need not be limited to
constant functions; certain kinds of biased functions can be
discriminated as well.

Unambiguous state discrimination is a procedure that is of
fundamental interest in quantum information theory.  Its only
application so far has been to quantum cryptography.  The results
presented here suggest that related methods can also serve as a
tool in the development of quantum algorithms.

This research was supported by the Office of Naval Research (Grant
No. N00014-92-J-1233), the National Science Foundation (Grant
No. PHY-0139692), the Hungarian Science Research Fund (Grant No. T
03061), a PSC-CUNY grant, and a CUNY collaborative grant.

\bibliographystyle{unsrt}

\end{document}